\documentclass[prl,twocolumn,amsmath,amssymb,superscriptaddress]{revtex4}
\usepackage{graphicx}
\usepackage{bm}

\begin{document}

\preprint{cond-mat/0504739}

\title{Tunneling density of states as a function of thickness in superconductor/ strong ferromagnet bilayers}

\author{S. Reymond}
\affiliation{Department of Applied Physics, Stanford University,
Stanford, CA 94305}
\author{P. SanGiorgio}
\email{psangior@stanford.edu}
\affiliation{Department of Physics, Stanford University, Stanford, CA 94305}
\author{M.R. Beasley}
\affiliation{Department of Applied Physics, Stanford University,
Stanford, CA 94305}
\author{J. Kim}
\affiliation{Center for Strongly Correlated Materials Research,
School of Physics, Seoul National University, Seoul 151-742,
KOREA}
\author{T. Kim}
\affiliation{Center for Strongly Correlated Materials Research,
School of Physics, Seoul National University, Seoul 151-742,
KOREA}
\author{K. Char}
\affiliation{Center for Strongly Correlated Materials Research,
School of Physics, Seoul National University, Seoul 151-742,
KOREA}

\date{\today}

\begin{abstract}
We have made an experimental study of the tunneling density of
states (DOS) in strong ferromagnetic thin films (CoFe) in proximity with
a thick superconducting film (Nb) as a function of $d_F$, the
ferromagnetic thickness.  Remarkably, we find that as $d_F$ increases,
the superconducting DOS exhibits a scaling behavior in which
the deviations from the normal-state conductance have a universal
shape that decreases exponentially in amplitude with characteristic
length $d^*\approx 0.4$~nm.  We do not see oscillations in the DOS
as a function of $d_F$, as expected from predictions based on
the Usadel equations, although an oscillation in $T_c(d_F)$ has
been seen in the same materials.

\end{abstract}


\maketitle

One of the incompletely solved problems in conventional
(noncuprate) superconductivity is the interaction between
superconductivity and magnetism. This issue arises most
prominently in the context of the so-called magnet superconductors
(e.g., CeCoIn$_5$) and in the superconductor/ferromagnet (SF)
proximity effect. One striking effect expected for a
superconductor in the presence of an exchange field is the
existence of spatial modulations of the superconducting pair wave
function (see for instance Ref.~\cite{DEMLER97}). These
oscillations occur in a new superconducting state where the center
of mass of pairs acquires a non-zero momentum. This state was
predicted 40 years ago and is known as the
Fulde-Ferrell-Larkin-Ovchinnikov (FFLO) state
\cite{FULDE64,LARKIN65}. In the case of the SF proximity effect,
this oscillating pair wave function is expected to exponentially
decay into the F layer. These oscillations, in turn, are predicted
to lead to oscillations in the critical temperature of SF bilayers
\cite{RADOVIC91,FOMINOV02}, inversions of the DOS
\cite{BUZDIN00,ZAREYAN01}, and changes in the sign of the
Josephson coupling in SFS sandwiches \cite{BUZDIN82} (creating a
so called $\pi$-junction), as the F layer thickness, $d_F$, is
varied. Even more exotic predictions include possible
odd-frequency triplet superconductivity \cite{BERGERET01}.

Indeed, a vast theoretical literature now exists regarding the SF
proximity effect, mostly concerning systems with a uniform
magnetization in the F layer in the dirty quasiclassical limit
(i.e. where all the characteristic lengths are larger than both
$\lambda_F$, the Fermi wavelength, and $\ell$, the mean free
path). In this situation, the superconducting properties can be
calculated using the Usadel equation \cite{USADEL70}. Still,
explicit predictions may differ depending on the importance of
scattering processes \cite{DEMLER97,CRETINON05} or on the boundary
conditions, which can be resistive \cite{FOMINOV02} or
magnetically active \cite{DOH04}.

Experimentally, phase sensitive measurements in some SFS
structures clearly demonstrate the existence of $\pi$-junctions
\cite{RYAZANOV01B,KONTOS02,GUICHARD02}. On the other hand,
critical temperature, $T_c$, measurements in SF structures have
shown a variety of behaviors as a function of $d_F$. Everything
from monotonic dependence to step-like features, small dips and
oscillations have been reported
\cite{STRUNK94,JIANG95,MUHGE98,LAZAR00,FOMINOV02,BOURGEOIS02}. It
has been pointed out that important parameters such as the size of
the exchange field, $E_{ex}$, and the boundary resistance,
$\gamma_B$, can evolve naturally as a function of
$d_F$~\cite{AARTS97}. What makes these results particulary hard to
interpret is that all of these changes typically take place within a few
nanometers, just like the expected oscillation of the
superconducting wave function inside the F material.

In short, the situation is complicated both theoretically and
experimentally, and no clear, comprehensive understanding has
emerged.  This suggests that some new experimental approaches that
probe superconductivity directly inside the F material  might be
helpful.

In this paper we present density of state spectroscopy studies
using tunneling junction located on the F side of
Nb/Co$_{0.6}$Fe$_{0.4}$ bilayers. CoFe is a strong ferromagnet
with a Curie temperature of approximately $1100$~K widely used in
magnetic tunnel junction devices due to the high quality of the
interface it makes with aluminum oxide, which is now the standard
choice for tunneling barriers. The critical temperatures of
similarly deposited Nb/CoFe bilayers, but with a thinner Nb layer
(18~nm), show an oscillatory behavior as a function of $d_F$: a
slight dip is noticeable before saturation at large $d_F$.  A
quantitative analysis of these data based on the Usadel equations
appears elsewhere \cite{KIM05}. Based on these data, we expect
that any inversion in the DOS that may arise in our samples would
occur at the same thickness as the dip in $T_c$, which is between
1 and 2~nm. Thus, we performed tunneling spectroscopy on samples
with thicknesses ranging from 0 to 4.5 nm in increments of 0.5~nm.

Our results are remarkable in that we find that the deviations of
the density of states from that of the normal state exhibit a
precise scaling in amplitude over many orders of magnitude as
$d_F$ is increased. No oscillations in the sign of these
deviations are observed.


\begin{figure}
\includegraphics{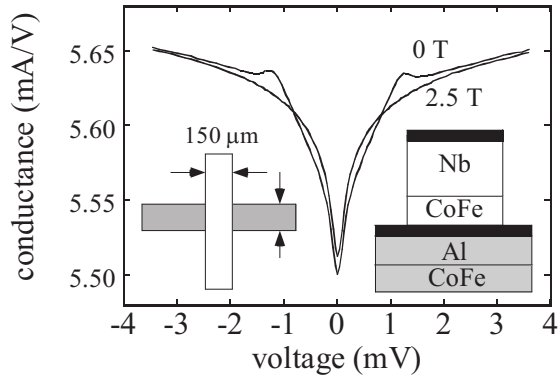}
\caption{\label{fig:norm} Raw conductance data taken at 0.5 K in a
Nb/CoFe/AlO$_x$/Al/CoFe junction at zero field and above the Nb
critical field. Here $d_F=2.5$ nm. Left inset: sample top view.
Right inset: junction cross section. The black areas represent Al
oxide layers.}
\end{figure}

Our junctions are deposited and patterned entirely {\it in situ}
in a DC magnetron sputtering system, described here \cite{KIM05}.
The full geometry of the tunneling structures is pictured in the
inset of Fig.~\ref{fig:norm}. A thin ($3$~nm) CoFe layer which
suppresses superconductivity in the Al electrode is deposited
first
--- we see no superconducting transition in the Al above $0.3$~K
--- then an atomic oxygen source is used to fully oxidize the
barrier to a thickness of $\approx 2$ nm. The stencil mask is then
changed and without breaking vacuum a CoFe layer and then a Nb
layer are immediately deposited, followed by a very thin Al cap to
protect the Nb layer from oxidation. In order to establish a
robust superconducting state in the S layer, we deposit
approximately $50$~nm of Nb, which is thick enough that there
should be only small changes in $T_c(d_F)$ \footnote{However the
Nb thickness was not tightly controlled and consequently our
samples display a non-systematic spread in $T_c$}. Typical
junction resistances are roughly $200~\Omega$, with a trend to
higher junction resistances with increased CoFe thickness.

Measurements shown here are taken at $0.5$~K using standard
lock-in techniques to measure the differential resistance;
variations as small as few parts in 10,000 could be distinguished.
This resolution is crucial since when $d_F$ exceeds 1~nm the
superconducting signal becomes very weak, reaching less than a
part per thousand at $d_F=3$~nm.

Figure \ref{fig:norm} shows typical conductance spectra taken at
0.5~K in zero-field and with a perpendicular field of 2.5~T, just
above the critical field, $H_{c2}$. The zero-field curve clearly
reveals the superconducting density of states, but it is
superposed on top of a zero-bias anomaly. The high field data show
the zero-bias anomaly on its own. This anomaly is a V-shaped
curve, centered at zero volts, with the conductance at $1$~mV
typically two percent larger than the zero-bias conductance. The
main experimental difficulty lies in extracting the
superconducting DOS from this zero-bias anomaly. To do this, we
divide the zero-field curve by the in-field curve. This
straightforward procedure alone was sufficient for the thicker
barriers, because no change in the normal conductance was seen
between perpendicular and parallel applied fields or as a function
of field strength. However for samples with thin CoFe layers (less
than $1.5$~nm), an additional zero-bias resistance peak increasing with
the applied field was seen above $H_{c2}$, similar to
previous studies~\cite{EIGLER04}. For these samples, the
zero-field normal-state background could be calculated by studying
the field-dependence of this feature and extrapolating it to
zero-field.

\begin{figure}
\includegraphics{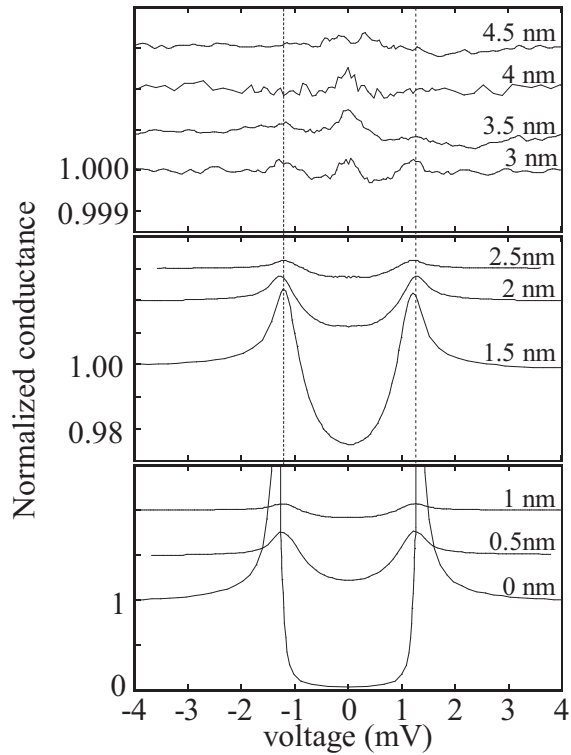}
\caption{\label{fig:didvall} Normalized conductances taken at
$0.5$~K for various CoFe thickness indicated inside. From the
bottom to the top plot the vertical scale is successively
amplified. The curves are shifted for clarity. The dashed line
shows that the peaks maximum remains unchanged from 0.5 to 3~nm.}
\end{figure}


Figure \ref{fig:didvall} shows the resulting curves for all
measured thicknesses. Note that in the top panel the scale is
amplified about a thousand times. Beginning at the bottom of the
figure, we see that for $d_F$ = 0 a clean DOS curve for Nb is
obtained, as expected. A BCS fit to this curve gives $1.3$~meV for
the gap energy. When the thinnest ($0.5$~nm) F layer is added to
the superconductor, the tunneling spectrum is abruptly altered
from a BCS-like shape to a much more smeared shape with
substantial conductance below the gap energy. As $d_F$ increases
further, the superconducting features in the tunneling conductance
are strongly attenuated. Above $2.0$~nm, a narrow zero-bias
conductance peak develops which is suppressed in fields greater
than $H_{c2}$; this suggests it is related to superconductivity,
but due to its weak signal-to-noise ratio, we do not focus on it
in the present discussion. When $d_F$ exceeds $3.5$ nm, all
recognizable features disappear and the normalized conductance is
equal to $1 \pm 10^{-4}$.

\begin{figure}
\includegraphics{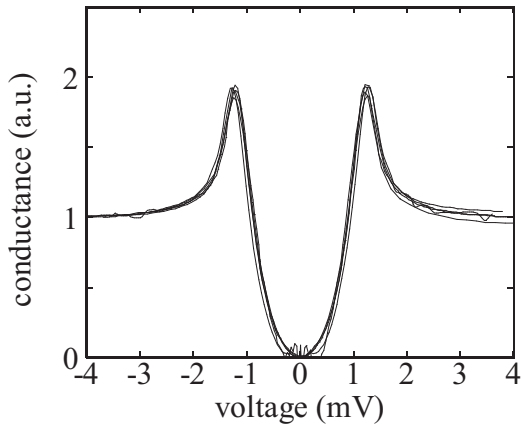}
\caption{\label{fig:stretch}Superposition of five scaled
conductance curves for $d_F=0.5$ to $2.5$ nm. The $d_F=2.5$~nm
curve is scaled by a factor of 500.}
\end{figure}

\begin{figure}
\includegraphics{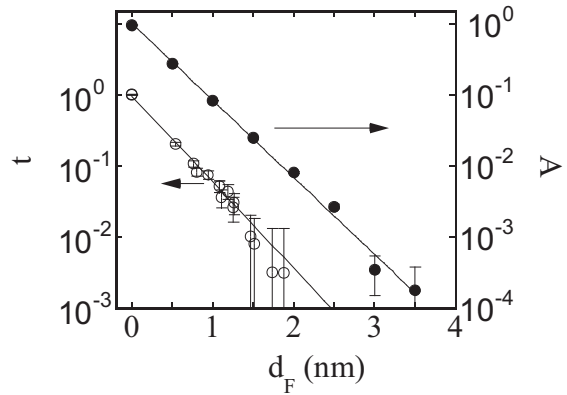}
\caption{\label{fig:zbc}Scaling factor, $A(d_F)$, and normalized
transition temperature, $t=(T_c-Min[T_c])/T_{c}(d_F=0)$, as a
function of $d_F$ for SF bilayers. For $t(d_F)$, only the initial
decay is plotted ($d_F<2$~nm). For both sets of data, the
characteristic decay length is $0.4$ nm.}
\end{figure}

The most striking observation, though, is that between 0.5 nm and
2.5 nm the spectra can be rescaled onto a single curve.  That is
to say,

\begin{equation}
\sigma(V,d_F)-1 = A(d_F)(N(V)-1)
\label{eq:factor}
\end{equation}
where $N(V)$ is a generic, thickness independent function and
$A(d_F)$ is a scaling coefficient defined so that $N(0)=0$. The
$N(V)$ curves derived from the data using this scaling procedure
are shown in Fig.~\ref{fig:stretch}. The overlap of the curves
demonstrates the generic nature of $N(V)$. $A(d_F)$ is plotted in
Fig.~\ref{fig:zbc} (full circles). The straight line is an
exponential fit with decay length $d^*=0.4$~nm. Note that for
$d_F>2$~nm, we disregard the narrow peak at $V=0$ when scaling
the curves. The fact that $A(d_F)$ extrapolates to 1 as $d_F$
tends to zero is not required by the scaling procedure. It has
the consequence that $N(V)$ can be interpreted physically as the
normalized tunneling DOS when some minimal thickness of F
material has been deposited below the S layer. Note also that
the zero-bias conductance is simply $1-A(d_F)$ and therefore
rises exponentially from zero as $d_F$ increases.

An important characteristic of the generic DOS $N(V)$ is that the
sum rule on the DOS is satisfied within one percent. This not only
justifies the procedure used to isolate the superconducting DOS,
even when it is much weaker than the ``normal'' background, it also
suggests that our tunnel barrier is not weakend by the addition
of the CoFe layer.  Such weakening would result in an excess current
within the gap, raising the total spectral area above
unity~\cite{BLONDER82}.

It is interesting to compare these DOS data with the $T_c(d_F)$
data reported earlier on related samples with a thinner Nb layer,
mentioned above. There are no evident oscillations in $A(d_F)$
near $d_F\approx 2$~nm, where an oscillation is seen in
$T_c(d_F)$. On the other hand, in Fig.~\ref{fig:zbc}, we compare
the initial drop in $T_c$ (open circles) with the scaling factor,
$A(d_F)$, derived from the DOS. Specifically, we plot the
normalized critical temperature, $t=(T_c-Min[T_c])/T_{c}(d_F=0)$.
Remarkably, both sets of data show an exponential decrease with
the same characteristic length scale.

The absence of oscillations in our DOS data contrasts with a previous
experimental study on a different material~\cite{KONTOS01}. In that study,
a weak ferromagnetic alloy (Pd$_{1-x}$Ni$_x$) was studied
at two different thicknesses and a robust inversion of the DOS was seen.
On the other hand, a subsequent study in which the Ni composition (and hence
$E_{ex}$) was varied at fixed $d_F$ showed scaling in the DOS similar
to that reported here~\cite{KONTOS04}.  It is natural that these
different approaches should yield similar results, as the relevant
magnetic length, $\xi_F$, is given by $\sqrt{\hbar D /2 E_{ex}}$, where
$D$ is the diffusion constant in the F layer. In addition, the authors
of Ref.~\cite{KONTOS04} note that for relatively thick ($d_F \gg \xi_F$)
samples such as theirs, scaling of the density of states with
exponentially decaying oscillations is predicted by the linearized
Usadel equations appropriate to that limit.  This is consistent with
their data~\cite{KONTOS05}, but it is clearly inconsistent with ours.

Almost any application of the Usadel formalism that starts with a large
exchange field will predict decaying oscillations in the DOS.  In order
to prevent these oscillations within this framework, one must include a
generic Abrikosov-Gorkov spin-breaking parameter, $\Gamma_{AG} \gtrsim E_{ex}$. 
Further, to obtain strictly exponential scaling of the DOS (in the
$d_F \gg \xi_F$ limit) one must take $\Gamma_{AG}$ to be much greater
than $E_{ex}$.  In this model, $d^* = \sqrt{\hbar D /4 \Gamma_{AG}}$,
which would neccessitate $\Gamma_{AG}=300$~meV to account for our data.

Although the above model gives a qualitative explanation for the two
key features of our data (the exponential nature of $A(d_F)$ and the
invariance of $N(V)$), it is not clear how accurately it corresponds to
the actual physical situation.  Our thinnest sample, with $d_F=0.5$~nm,
for instance, is certainly not in the semi-infinite limit, $d_F \gg d^*$.
Thus, we would expect finite-size effects, such as a decrease of the gap width,
which would prevent it from following the predicted scaling relationship.  Further,
one must wonder about the origin of $\Gamma_{AG}$ and the seeming
unimportance of $E_{ex}$.  Previous studies on thin CoFe films have confirmed
that near-bulk ferromagnetic ordering exists in films as thin as
1~nm~\cite{WANG98}, and further, it is reasonable to suggest that given the
constant shape of the DOS, none of the important physical parameters are changing
significantly over the thicknesses we examine.  Therefore, we conclude
that a constant exchange field is present in all of our samples with a CoFe layer.
Another possible mechanism for washing out the expected oscillations could
be lateral variations in $d_F$~\cite{HALTERMAN02}, but we note that these
variations would have to be vary large ($\delta d_F \sim \xi_F$), which seems
unlikely.  Finally, we note that the Usadel equation is only strictly valid
in the diffusive limit, wherein all relevant length scales are larger than the
mean free path.  From resistivity data, we estimate $\ell = 1.0$~nm, which is
somewhat larger than $d^* = 0.4$~nm, which is comparable in size to
$\lambda_F = 0.3$~nm. In light of the totality of the considerations above,
we come to the position that no reasonable application of the conventional
Usadel theory or purely materials problem can account for our results.


In summary, we have measured the tunneling DOS on the F side of SF
bilayers and have found a sharp change in the shape of the
conductance curves between $d_F=0$ and $d_F=0.5$~nm and that for
all $d_F > 0$, the shape of the conductance curves is universal.
The only dependence on $d_F$ is given by an exponential decrease
in the magnitude of the superconducting signal with characteristic
length $d^* = 0.4$ nm. We also note a similar exponential decrease
in $T_c(d_F)$ in related samples. Finally, we note that we have been
unable to reconcile our results with the conventional Usadel equation;
thus, new theoretical considerations appear necessary.

We thank L. Litvak and Y. Bazaliy for stimulating discussions. We
also thank T. Kontos and O. Valls for helpful communications. The
authors acknowledge the support of U.S. DOE, NSF, Swiss National
Science Foundation, and KOSEF through CSCMR.

\end{document}